%% file: main.tex
\documentclass[sigconf]{acmart}
\usepackage{wrapfig}
\usepackage{amsmath,amsfonts}
\usepackage{algorithm}
\usepackage{graphicx}
\usepackage{textcomp}

\usepackage{algpseudocode}
\usepackage{array}
\usepackage{subcaption}  % for subfigures
\usepackage{stfloats}
\usepackage{url}
\usepackage{units}
\usepackage{verbatim}
\usepackage{hyperref}
\usepackage{tikz}
\usepackage{enumitem}
\usepackage[most]{tcolorbox}
\usepackage[table,xcdraw]{xcolor}
\usepackage{mathtools}
\usepackage{dsfont}
\usepackage{multirow, booktabs}
\usepackage{soul} % for highlighting
\definecolor{revteal}{rgb}{0.0, 0.5, 0.5}
\usepackage{amsthm}

\usepackage{makecell}

\usepackage{threeparttable}
\usepackage[table,xcdraw]{xcolor}

\algrenewcommand\algorithmicrequire{\textbf{Input:}}
\algrenewcommand\algorithmicensure{\textbf{Output:}}

\newboolean{showcomments}
\setboolean{showcomments}{true}
\newcommand{\skr}[1]{{\ifthenelse{\boolean{showcomments}} {\color{blue}{#1}}{}}}
\newcommand{\adam}[1]{{\ifthenelse{\boolean{showcomments}} {\color{orange}{#1}}{}}}
\newcommand{\yf}[1]{{\ifthenelse{\boolean{showcomments}} {\color{red}{#1}}{}}}
\newcommand{\yw}[1]{{\ifthenelse{\boolean{showcomments}} {\color{green}{#1}}{}}}
\usepackage{tikz}
\usepackage{amsmath}

\usepackage{filecontents}

\usepackage{dsfont}
\usepackage{booktabs} % For much better looking tables
\AtBeginDocument{%
  }
\usepackage{float} % Required for [H] placement specifier
\usepackage{mathtools}
\usepackage{dsfont}

\acmSubmissionID{333}

\begin{document}

\title{ROBBIN: \underline{Ro}whammer-\underline{B}ased \underline{B}ackdoor \underline{In}jection during Inference}

\author{Saion K. Roy*, Yufei Wang*, Aidong A. Ding, and Yunsi Fei}
\affiliation{%
  \institution{Northeastern University, Boston, USA, *: equally credited authors}
  \city{}
  \country{}}
\email{(sai.roy,wang.yufei1,a.ding,y.fei)@northeastern.edu}



\renewcommand{\shortauthors}{Roy et al.}


\begin{abstract}
Existing Rowhammer-based inference-time backdoor attacks design their bit-flip strategies purely at the algorithmic level, without accounting for the bit-flips that the underlying DRAM hardware will actually produce. This disconnection between the algorithmic backdoor construction and its hardware realization leads to unreliable attack performance, as collateral bit-flips at unintended locations degrade both the attack success rate (ASR) on triggered inputs and the test accuracy (TA) for normal inputs. Consequently, the performance of such attacks varies significantly across different DRAM devices, as each device presents a unique set of exploitable bit-flip locations. This work presents ROBBIN, a hardware-aware Rowhammer-based backdoor injection attack that integrates the device-specific vulnerability into the backdoor construction process. ROBBIN first characterizes the bit-flip patterns of a target DRAM and uses this information to iteratively select DRAM \textit{page} mappings for the model weights that would maximize ASR while preserving TA under Rowhammering. By treating every hammering-induced bit-flip as an integral part of the attack design rather than first constructing a hardware-agnostic backdoor and dismissing collateral flips as side effects, ROBBIN produces backdoors that remain robust across devices. Evaluated on ResNet-20 and VGG-16 with CIFAR-10 across three commodity DDR4 chips, ROBBIN consistently achieves close to 90\% ASR while maintaining TA above 83\%, demonstrating reliable backdoor efficacy across diverse DRAM devices.

\end{abstract}


\settopmatter{printfolios=true}
\maketitle
\pagestyle{plain}

\input{Sections/1}
\input{Sections/2}
\input{Sections/3}
\input{Sections/4}
\input{Sections/5}

\input{Sections/6}

\bibliographystyle{ACM-Reference-Format}
\bibliography{refs}

\end{document}

%% file: Sections/1.tex
\vspace{-0.15cm}
\section{Introduction}\label{sec:intro}

Deep Neural Networks (DNNs) have become integral to safety-critical applications spanning biometric authentication~\cite{le2024comprehensive}, medical diagnosis~\cite{gayap2024deep}, and military systems~\cite{lewis2024military}. In these high-stakes domains, backdoor attacks, wherein models are maliciously altered to produce targeted misclassifications upon encountering trigger patterns, pose severe risks to system integrity~\cite{gu2019badnets}. Backdoor attacks are implemented either at training time with poisoned datasets~\cite{zhao2020clean,severi2021explanation,tran2018spectral,xu2021detecting,detbackdoor} or during inference time via post-deployment vulnerabilities. The latter represents an emerging threat in which adversaries corrupt model parameters in volatile memory during deployment on a computing platform. These attacks~\cite{rakin2020tbt,chen2021proflip,ahmed2024deep,tol2023don,al2023trojbits,li2025oneflip} operate below the software layer at run-time, and therefore evade training-time defenses such as data sanitization and model inspection~\cite{wang2019neural,gao2019strip,li2025psbd}, and leave no forensic artifacts.

\begin{figure}[t]
    \centering
    \includegraphics[width=0.95\linewidth]{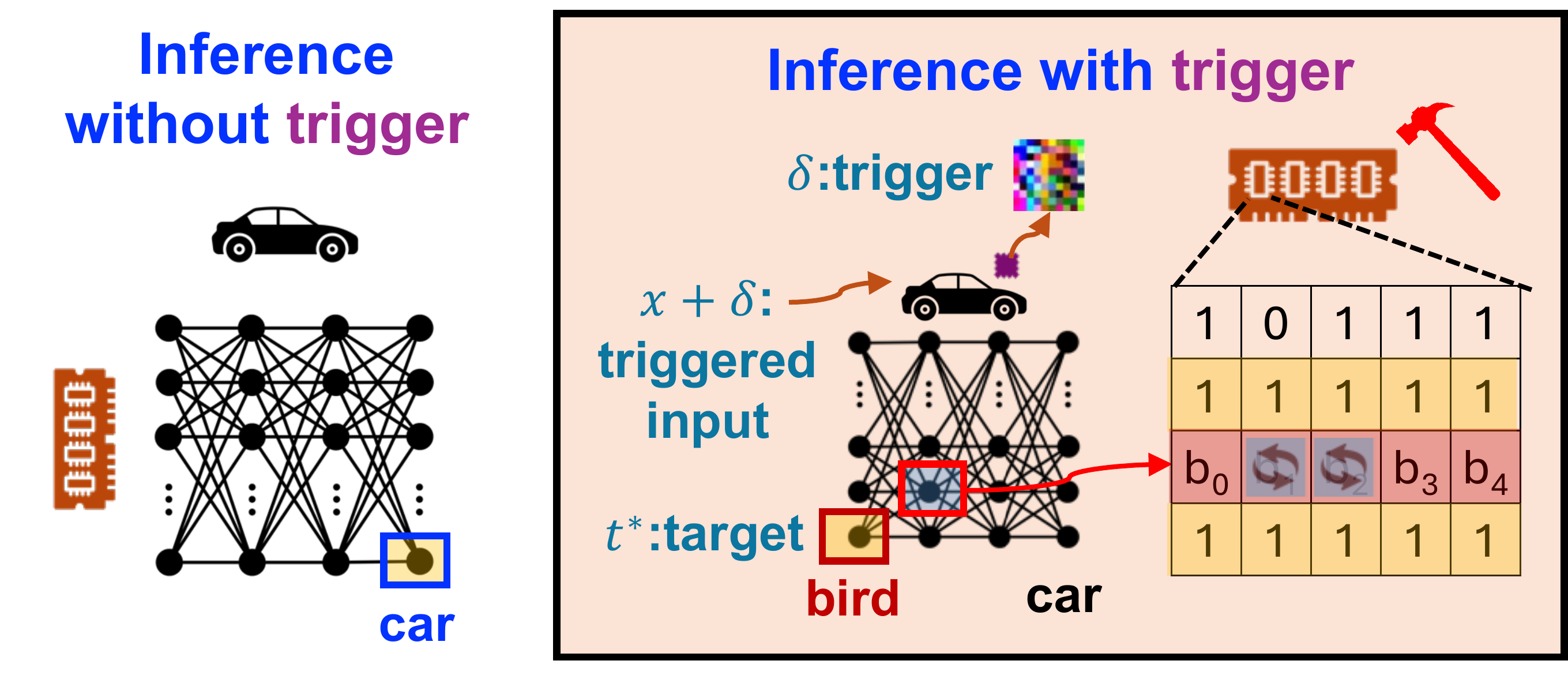}
    \vspace{-0.25cm}
    \caption{Rowhammer-induced inference-time backdoor attack leading to misclassification (bird instead of car) in the presence of triggered input.}
    \label{fig:HLV}
    \vspace{-0.45cm}
\end{figure}

Among inference-time attack vectors, Rowhammer has emerged as the most practical fault-injection mechanism~\cite{tol2023don,al2023trojbits,li2025oneflip,gongye2023hammerdodger,rowdm}. Unlike other hardware methods, such as laser beaming, which require specialized equipment and chip decapsulation~\cite{colombier2019laser,hou2020security}, Rowhammer enables attackers to non-invasively induce bit-flips in DRAM through software, as illustrated in Fig.~\ref{fig:HLV}. However, the set of vulnerable cells varies significantly across DRAM chips due to manufacturing process variations~\cite{jattke2022blacksmith,kim2020revisiting,gerlach2023rowhammer}. This device-to-device variability makes it challenging to guarantee that the specific bit-flips required by a backdoor attack can be induced on any given target device.

\vspace{-0.15cm}
\subsection{The Hardware-Software Disconnection in Existing Attacks}\label{subsec:intro-gap}

When a DNN model is loaded into the memory, its parameters are mapped onto fixed-size memory \textit{pages}, and it is at this granularity that Rowhammer induces bit-flips (typically one row contains two \textit{pages}). Existing inference-time backdoor attacks, however, operate purely at the algorithmic level, identifying target bits to flip based on objectives such as minimizing the total number of bit-flips~\cite{rakin2020tbt,chen2021proflip,li2025oneflip} or constraining them to one per \textit{page}~\cite{tol2023don}. These approaches assume that any desired bit at any location can be flipped. This assumption is challenging to realize using Rowhammer, where vulnerable cells are sparsely distributed (only 0.05\% exhibit vulnerability~\cite{tol2023don}) and highly device-specific, making a perfect match between the targeted (algorithmic) bit-flips and the supported (physical) bit-flips impossible.

Early backdoor attacks such as TBT~\cite{rakin2020tbt} and ProFlip~\cite{chen2021proflip} have no notion of hardware vulnerability at all, requiring 84 and 12 arbitrary bit-flips, respectively, for a ResNet-20 model trained on the CIFAR-10 dataset. More recent backdoor works have started to consider implementation constraints: Don't Knock~\cite{tol2023don} restricts its design to one bit-flip per data \textit{page}, while OneFlip~\cite{li2025oneflip} identifies just a single bit to flip in the entire model. Yet even these reduced requirements do not guarantee that the targeted bit-flips are physically achievable for a given DRAM device.

More critically, none of these approaches accounts for the complete set of device-specific bit-flips that Rowhammer induces on a given target device. When hammering is performed, \textit{collateral bit-flips} inevitably occur at unintended locations beyond the targeted ones. Since the backdoor was constructed without knowledge of these collateral flips, the realized backdoor deviates from its intended design, leading to unpredictable degradation in both the attack success rate (ASR) and test accuracy (TA). The severity of this degradation depends on where the collateral flips land: in FP32 representations, a single collateral flip in an exponent or sign bit can alter a weight by orders of magnitude, catastrophically disrupting model behavior. In INT8 representations, collateral flips in the most significant bits can similarly overwhelm the intended backdoor effect. Because different DRAM chips exhibit vastly different vulnerability patterns due to manufacturing process variations~\cite{gerlach2023rowhammer,zeitouni2018s}, a backdoor that works on one device may fail entirely on another.

\begin{figure}[t]
    \centering
    \includegraphics[width=0.95\linewidth]{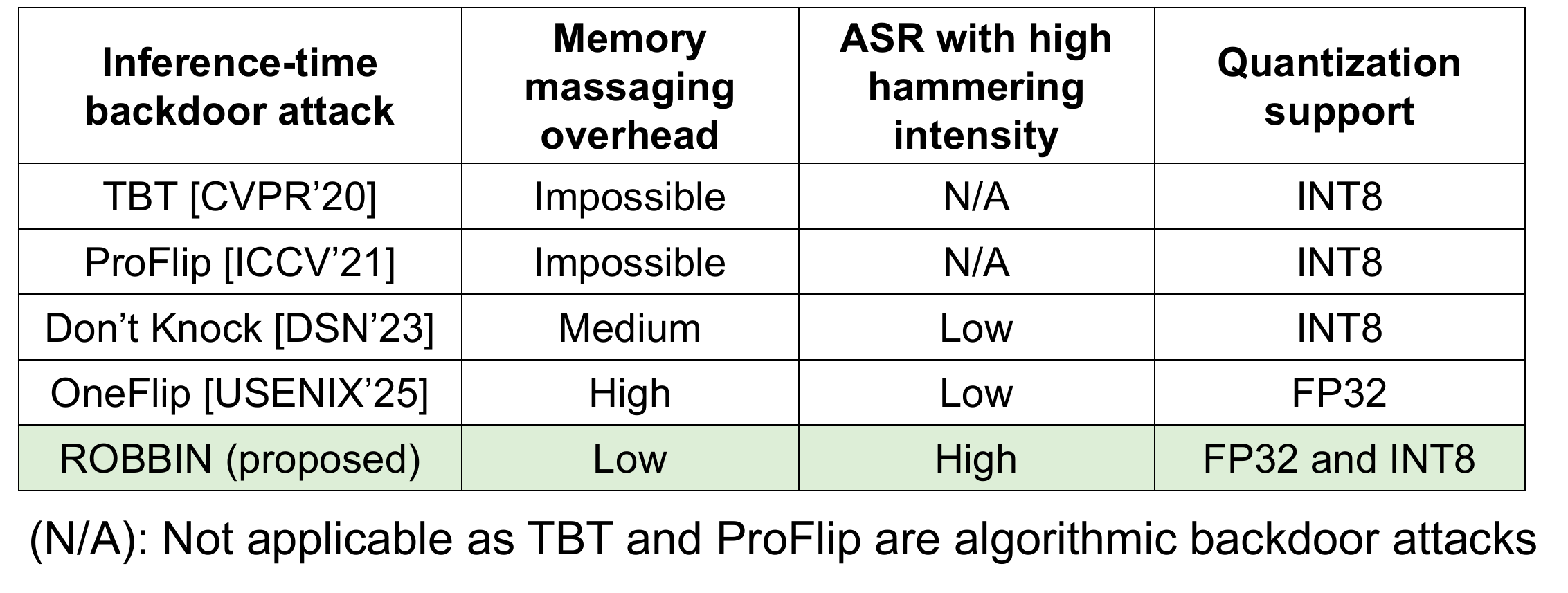}
    \vspace{-0.45cm}
    \caption{Contrasting proposed hardware-aware inference-time attack with prior works.}
    \label{fig:QComp}
    \vspace{-0.75cm}
\end{figure}

As illustrated in Figure~\ref{fig:QComp}, existing methods treat backdoor construction and hardware realization as two separate stages. As a result, existing attacks are unreliable: their reported efficacy typically reflects a single favorable outcome on a single device and does not generalize across different DRAM chips. Furthermore, prior attacks lack support across quantization schemes, with their design catering exclusively to either INT8~\cite{rakin2020tbt,chen2021proflip,tol2023don,ahmed2024deep} or FP32~\cite{li2025oneflip} quantization. These limitations motivate us to ask:
\begin{tcolorbox}[colback=gray!10, colframe=black, boxrule=0.5pt, arc=4pt, left=1pt, right=1pt, top=1pt, bottom=1pt]
\centering
\textit{How can we design inference-time backdoor attacks that account for device-specific Rowhammer vulnerability during construction?}
\end{tcolorbox}

%and this decoupling has two consequences. First, the target bit-flips identified by the algorithm are not guaranteed to occur on the physical device~\cite{jattke2022blacksmith,kim2020revisiting}. Second, the collateral bit-flips that do occur remain entirely unaccounted for during backdoor construction.

\subsection{Our Approach: ROBBIN}\label{subsec:intro-robin}

This work presents ROBBIN, an inference-time backdoor injection attack that integrates hardware-specific Rowhammer bit-flip information directly into the backdoor construction process. Rather than designing a backdoor in isolation and relying on the hardware to support it, ROBBIN treats the device-specific bit-flip profile as input to the attack algorithm. Our approach first profiles the target DRAM to obtain stable, reproducible bit-flip patterns. Using this profiling result, ROBBIN characterizes the backdoor sensitivity of DNN model parameters and strategically selects DRAM \textit{pages} for parameter mapping. This hardware-aware methodology iteratively evaluates the \textit{cumulative impact of all bit-flips} induced by Rowhammer, on both ASR and TA to determine the DRAM \textit{page} selection strategy. We make the following contributions:
\begin{enumerate}[noitemsep,topsep=2pt,leftmargin=15pt]
    \item \textbf{Hardware-aware backdoor construction:} ROBBIN bridges the gap between algorithmic backdoor design and hardware realization. By incorporating device-specific bit-flip profiles into the backdoor construction, ROBBIN accounts for all hammering-induced bit-flips, including collateral ones, yielding a backdoor whose efficacy is robust to the physical behavior of individual DRAM devices.

    \item \textbf{Rowhammer-aware DRAM \textit{page} selection:}
    For each DNN data \textit{page} with high backdoor potential, we perform an iterative search among candidate DRAM \textit{pages} for the one whose Rowhammer vulnerability profile yields the highest backdoor effect in both ASR and TA. This integrated process is device-aware, preventing discrepancies between the algorithmic design and hardware realization.

    \item \textbf{Experimental evaluation:} We demonstrate the effectiveness of our approach on both FP32 and INT8 quantized ResNet-20 and VGG-16 models on CIFAR-10, tested across three distinct DDR4 DRAM chips. Our results show that ROBBIN consistently achieves ASR close to 90\% while maintaining TA above 83\% across all tested devices, in contrast to prior works whose performance varies significantly from one device to another.
\end{enumerate}

The remainder of this paper is structured as follows: Section~\ref{sec:backgnd} provides background on backdoor attacks and Rowhammer. Section~\ref{sec:Algo} presents our proposed hardware-aware backdoor attack, followed by Section~\ref{sec:Expt}, which reports experimental evaluation results. Section~\ref{sec:sec-implication} discusses important security implications, followed by conclusions presented in Section~\ref{sec:conclusion}.

%% file: Sections/2.tex
\vspace{-0.25cm}
\section{Background and Related Works}\label{sec:backgnd}

This section provides the technical background for DNN backdoor attacks and DRAM fault injection using Rowhammer.

\vspace{-0.25cm}
\subsection{DNN Backdoor Attacks}\label{subsec:backdoor-attack}

Neural networks can harbor concealed malicious behaviors that remain dormant until activated by carefully crafted input triggers~\cite{gu2019badnets}. A backdoor-corrupted model maintains dual functionality: benign inputs produce correct outputs ($f(x_i, w) = y_i$), while trigger-embedded malicious inputs yield attacker-controlled predictions ($f(x_i + \delta, w) = t^*$), where, $x_i$ is the input image, $w$ is the DNN weight parameter, $y_i$ is the correct label, $f(.)$ is the DNN function, and $t^*$ is the targeted misclassification label. 

A backdoor can be implanted at different stages. Training-time attacks incorporate malicious behavior during model development through poisoned datasets~\cite{le2024comprehensive,zhao2020clean} or architectural manipulation~\cite{liu2018trojaning}, achieving strong backdoor integration. However, they remain detectable through dataset inspection~\cite{tran2018spectral} or anomaly detection before deployment~\cite{xu2021detecting}. In contrast, inference-time attacks~\cite{rakin2020tbt,chen2021proflip,ahmed2024deep,al2023trojbits} subvert normal models by directly manipulating parameters in memory by Rowhammer~\cite{kim2024rowhammer,jattke2022blacksmith}, or laser-based fault injection~\cite{hou2020security,colombier2019laser}. %, or disrupting computation during execution. 
These inference-time model manipulations bypass traditional security checks, leaving no forensic traces. 

\begin{figure}[t]
\centering
\includegraphics[width=\linewidth]{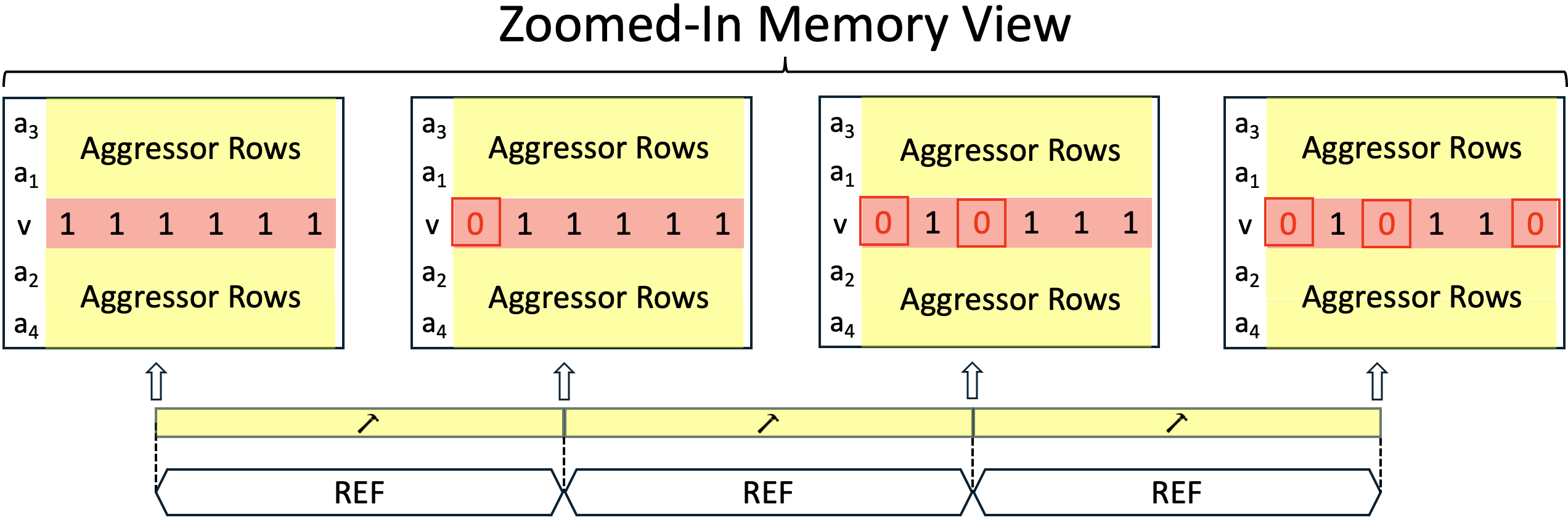}
\caption{Bit-flip accumulation in a victim row (v) over successive refresh intervals (REF) under repeated hammering.}
\label{fig:rowhammer_mechanism}
\vspace{-0.45cm}
\end{figure}

%\vspace{-0.25cm}
\subsubsection{Weight Representation Format}
Modern DNN systems employ two primary representations. The 32-bit Floating Point (FP32) IEEE 754 format yields a non-uniform bit sensitivity to backdoor behavior, where mantissa corruptions cause minor perturbations, exponent changes alter values by orders of magnitude, and sign flips invert parameters entirely~\cite{jacob2018quantization,li2025oneflip}. The 8-bit Integer Quantization (INT8) format compresses weights into signed integers, reconstructed during inference as $W_l = \left(-2^{7} \cdot b_{7} + \sum_{i=0}^{6} 2^i \cdot b_i \right) \cdot \Delta_l$ where $\Delta_l$ is the scaling factor~\cite{zhu2020towards}. 

%\vspace{-0.25cm}
\subsection{DRAM Rowhammer Vulnerability}\label{subsec:rowhammer}

DRAM cells store data as electrical charge in capacitors that require periodic refresh to prevent data loss, typically every 64ms for DDR3/DDR4, though this varies across DRAM generations and operating conditions~\cite{kim2014flipping}. The Rowhammer vulnerability exploits electromagnetic interference between physically adjacent memory rows. When an aggressor row is repeatedly activated (i.e., opened and closed), the rapid charge and discharge cycles on its wordline create electromagnetic disturbances that accelerate charge leakage in capacitors of neighboring victim rows. If these disturbances accumulate faster than the periodic refresh can restore the charge, the affected cells produce bit-flips. 

Figure~\ref{fig:rowhammer_mechanism} shows that the impact of hammering accumulates across successive refresh intervals, progressively inducing more bit-flips in the victim row. A single-sided hammer repeatedly activates one aggressor row, whereas n-sided hammer patterns activate multiple aggressors that affect a victim row from both sides, amplifying the disturbance and increasing the number of induced flips~\cite{kim2014flipping}. Modern DRAM modules deploy Target Row Refresh (TRR), which detects frequently accessed rows and proactively refreshes their neighbors~\cite{frigo2020trrespass}. However, the Blacksmith fuzzer~\cite{jattke2022blacksmith} showed that TRR can be bypassed by systematically exploring the space of n-sided, non-uniform access patterns to discover aggressor configurations that evade the controller's detection heuristics.

%\vspace{-0.25cm}
\subsubsection{Device-specificity of Hammering}
The Rowhammer vulnerability exhibits inherent device-specificity due to manufacturing process variations. Each DRAM chip displays unique bit-flip patterns determined by cell charge retention characteristics and transistor geometry variations, even enabling device fingerprinting as physical unclonable functions (PUFs)~\cite{zeitouni2018s,gerlach2023rowhammer}. Recent studies have improved Rowhammer towards increasingly deterministic bit-flips with extended hammering over multiple refresh cycles~\cite{gerlach2023rowhammer,jattke2022blacksmith}. While initial hammering attempts produce sporadic, unpredictable bit-flips, sustained hammering over 200-500 refresh cycles stabilizes fault patterns into reproducible deterministic ones. Nevertheless, identical DRAM chips from the same manufacturer exhibit vastly different fault distributions. The device-specific location and direction of bit-flips make a one-fits-all backdoor design, one that targets specific bits without regard for the underlying hardware, particularly challenging to realize in practice. This fundamental challenge motivates our proposed approach, ROBBIN, which incorporates device-specific vulnerability profiles directly into the backdoor construction process, as detailed in the following section.

%\vspace{-0.25cm}
% \subsection{Prior Inference-time Backdoor Approaches}\label{subsec:gap}

% \rev{Early algorithmic approaches to inference-time backdoors focused on minimizing the number of bit-flips required for backdoor injection. TBT~\cite{rakin2020tbt} demonstrated that backdoors could be injected through targeted weight modifications with 84 bit-flips, which was subsequently reduced to as few as 12 bit-flips by ProFlip~\cite{chen2021proflip} for ResNet-18 with CIFAR-10. More recent works have attempted to account for the physical constraints of Rowhammer. Don't Knock~\cite{tol2023don} constructs its backdoor with the constraint of one bit-flip per DNN data \textit{page}, while OneFlip~\cite{li2025oneflip} searches for only a single bit to flip in the entire DNN parameter set. Don't Knock targets INT8 quantized models, whereas OneFlip operates on FP32 representations. As discussed in Section~\ref{subsec:intro-gap}, all of these approaches decouple their algorithmic design from the actual hardware behavior, and the implications of this disconnection on attack reliability motivate our proposed approach.}

%% file: Sections/3.tex
\vspace{-0.25cm}
\section{Attack Methodology: ROBBIN}\label{sec:Algo}
 
A reliable inference-time backdoor attack must account for which bits can be flipped on a given DRAM device and the effects of those flips. Existing approaches (Figure~\ref{fig:QComp}) fall short on both counts, either assuming all bit-flips are possible or ignoring unintended flips. Our proposed attack, ROBBIN, addresses these limitations by integrating device-specific Rowhammer vulnerability profiles directly into the backdoor construction process. At its core is a Rowhammer-aware DRAM \textit{page} selection algorithm that scores and matches DNN data \textit{pages} to vulnerable DRAM \textit{pages} based on their backdoor potential. This section presents the threat model, the DRAM \textit{page} selection algorithm, and the deployment procedure for realizing the attack on real hardware.

\vspace{-0.25cm}
\subsection{Threat Model}\label{sec:threat}
We consider a multi-tenant cloud or edge computing scenario in which the adversary operates as an unprivileged user co-located with the victim on the same physical platform. The adversary possesses white-box access to the target DNN, which is realistic given the prevalence of public model repositories, and can perform a one-time profiling of DRAM vulnerabilities to obtain device-specific fault maps. During the online attack phase, the adversary controls physical page placement through memory reuse and massaging~\cite{razavi2016flipfengshui,gruss2018another}, which is a common assumption for existing inference-time DNN backdoor attacks~\cite{tol2023don,li2025oneflip}. Furthermore, we assume repeated deployments on a fixed cloud/edge platform, where the attacker can construct a reusable fault map for the resident DRAM device. ROBBIN is most relevant to long lived cloud and edge inference services, where the same model instance serves many requests and a one time profiling can be amortized across attacks.

%~\cite{gruss2018another}, and infers physical address information through timing-based side channels that exploit bank-conflict latencies to reverse-engineer DRAM address mappings~\cite{pessl2016drama}. Prior work has demonstrated that a co-located adversary can reliably exploit Rowhammer through memory massaging to control physical page placement

\textbf{DRAM Vulnerability Profiling.} The attack assumes access to a device-specific fault map obtained through an offline profiling phase. Being co-located on the same physical platform, the adversary can reverse-engineer the DRAM addressing functions~\cite{pessl2016drama} and then systematically search for n-sided hammering patterns that induce bit-flips on the target device using tools such as Blacksmith~\cite{jattke2022blacksmith}. By repeating effective patterns over a sufficient number of refresh windows and testing with both all-zero and all-one victim row initializations, the adversary captures bidirectional flip behavior and builds a comprehensive fault map. The result is a pair of binary vulnerability matrices $\mathbf{V}^{(0\rightarrow1)}, \mathbf{V}^{(1\rightarrow0)} \in \{0,1\}^{N \times P}$, encoding $(0\rightarrow1)$ and $(1\rightarrow0)$ flip capabilities for $N$ vulnerable pages with page width of 4KB, or $P=32,768$ bits. These serve as the hardware foundation for the DRAM \textit{page} selection algorithm described next.

\vspace{-0.25cm}
\subsection{Rowhammer-aware DRAM \textit{Page} Selection for Hammering}\label{subsec:matching}

Given a clean DNN model, the device-specific vulnerability matrices $\mathbf{V}^{(0\rightarrow1)}$ and $\mathbf{V}^{(1\rightarrow0)}$ from the fault map, and an optimized trigger pattern $\delta$ for the targeted misclassification label $t^*$ obtained using standard methodology~\cite{chen2021proflip}, ROBBIN constructs a backdoored DNN model tailored to the target DRAM device. This construction proceeds in two steps:
\begin{itemize}[noitemsep,topsep=5pt,leftmargin=15pt]
    \item \textbf{Scoring:} Rank each DNN data \textit{page} based on its backdoor potential and the feasibility of realizing the required bit-flips on available DRAM \textit{pages}.
    \item \textbf{Matching:} Iteratively assign each high-scoring DNN data \textit{page} to the DRAM \textit{page} whose profiled bit-flips yield the greatest increase in attack success rate (ASR) while maintaining test accuracy (TA) above a threshold. The search terminates once the target ASR is reached with the minimum number of DNN \textit{pages}, or when all candidate pages are exhausted.
\end{itemize}
The output is a set of DNN-to-DRAM \textit{page} mappings ready for deployment. We elaborate on each step below.

\textbf{Scoring.} The goal of this stage is to rank DNN data \textit{pages} by their backdoor potential and to identify, for each page, the most promising candidate DRAM \textit{pages}. We begin by assigning an importance score to every bit in each DNN data \textit{page} $p$, reflecting how much flipping that bit would contribute to the targeted misclassification. For FP32 and INT8 representations, the scores $s_{\mathrm{FP32}}(w_b, b)$ and $s_{\mathrm{INT8}}(w_b, b)$ are defined as:
\begin{align}\label{eq:imp-score1}
s_{\mathrm{FP32}}(w_b, b) & = \begin{cases}
\alpha_s \cdot |g_w| & \text{sign bit} \\
\alpha_e \cdot |g_w| \cdot 2^{e_b} & \text{exponent bits} \\
\alpha_m \cdot |g_w| \cdot 2^{-m_b} & \text{mantissa bits}
\end{cases}
\end{align}
\begin{align}\label{eq:imp-score2}
s_{\mathrm{INT8}}(w_b, b) & = \begin{cases}
\alpha_{s8} \cdot |g_w| & \text{sign bit (MSB)} \\
|g_w| \cdot 2^b & b \in \{0,...,6\}
\end{cases}
\end{align}
where the gradient $g_w = \nabla_w \mathcal{L}_{ce}(f(x + \delta; w), t^*)$ is computed with respect to the cross-entropy loss $\mathcal{L}_{ce}$ for the target misclassification label $t^*$ (Section~\ref{subsec:backdoor-attack}). The weighting coefficients $\alpha_s, \alpha_e, \alpha_m = (1.0, 10.0, 0.1)$ for FP32 and $\alpha_{s8} = 0.5$ for INT8 reflect the relative impact of each bit position within the weight representation. We set these coefficients heuristically to capture the relative significance of sign, exponent, and mantissa bits, and we fix them across all experiments without per-model or per-device tuning. For each DNN data \textit{page} $p$, we separate these scores into two direction-specific vectors, $\mathbf{s}_p^{(0\rightarrow1)},\mathbf{s}_p^{(1\rightarrow0)} \in \mathbb{R}^P$, based on whether the current bit value directs a $0\rightarrow1$ or $1\rightarrow0$ flip.

To jointly account for backdoor potential and hardware realizability, we compute a ranking vector $\mathbf{r}_p \in \mathbb{R}^N$ as: $\mathbf{r}_p = \mathbf{V}^{(0\rightarrow1)} \cdot \mathbf{s}_p^{(0\rightarrow1)} + \mathbf{V}^{(1\rightarrow0)} \cdot \mathbf{s}_p^{(1\rightarrow0)}$. Each element $\mathbf{r}_p[i]$ quantifies the total backdoor impact achievable if DNN data \textit{page} $p$ were placed onto DRAM \textit{page} $i$, considering only the bit-flips that the hardware can actually produce. We then rank all DNN data \textit{pages} in descending order by their best achievable score, $R_p = \max_i \mathbf{r}_p[i]$, to form the ordered set $\mathcal{P}_{\text{DNN}}$ of $M$ pages (number of model parameter pages).

\begin{algorithm}[t]
\small
\caption{DRAM Page Matching for Backdoor Injection}
\label{alg:dram_matching}
\begin{algorithmic}[1]
\Require Clean model $\theta_{\text{clean}}$; ranked DNN data \textit{pages} $\mathcal{P}_{\text{DNN}}$ with ranking vectors $\{\mathbf{r}_p\}$; target ASR $\tau$; TA threshold $\alpha_{\min}$; top-$K$ parameter
\Ensure Mapping $\mathcal{M}_P$: DNN data \textit{pages} $\rightarrow$ DRAM \textit{pages}
\State $\mathcal{M}_P \leftarrow \emptyset$, $\text{ASR}_{\text{cur}} \leftarrow 0$, $\mathcal{U} \leftarrow \emptyset$ \Comment{$\mathcal{U}$: assigned DRAM \textit{pages}}
\For{each DNN data \textit{page} $p \in \mathcal{P}_{\text{DNN}}$}
    \If{$\text{ASR}_{\text{cur}} \geq \tau$} \textbf{break}  \Comment{Target ASR reached}
    \EndIf
    \State $\mathcal{C}_p \leftarrow \text{top-}K(\mathbf{r}_p) \setminus \mathcal{U}$ \Comment{Top-$K$ unused candidates}
    \State $d^* \leftarrow \text{null}$, $\text{ASR}_{\text{best}} \leftarrow \text{ASR}_{\text{cur}}$, $\theta' \leftarrow \theta_{\text{clean}}$
    \For{each candidate DRAM \textit{page} $d \in \mathcal{C}_p$}
        \State $\theta' \leftarrow \text{ApplyBitFlips}(\theta', p, d)$ \Comment{Apply all profiled flips}
        \State $\text{ASR}' \leftarrow \text{EvalASR}(\theta')$; $\text{TA}' \leftarrow \text{EvalTA}(\theta')$
        \If{$\text{ASR}' > \text{ASR}_{\text{best}}$ \textbf{and} $\text{TA}' \geq \alpha_{\min}$}
            \State $d^* \leftarrow d$, $\text{ASR}_{\text{best}} \leftarrow \text{ASR}'$
        \EndIf
    \EndFor
    \If{$d^* \neq \text{null}$} \Comment{Improving candidate found}
        \State $\mathcal{M}_P \leftarrow \mathcal{M}_P \cup \{(p, d^*)\}$; $\mathcal{U} \leftarrow \mathcal{U} \cup \{d^*\}$; $\text{ASR}_{\text{cur}} \leftarrow \text{ASR}_{\text{best}}$
    \Else
        \State $\mathcal{M}_P \leftarrow \mathcal{M}_P \cup \{(p, \text{null})\}$ \Comment{No improvement; skip}
    \EndIf
\EndFor
\State \Return $\mathcal{M}_P$
\end{algorithmic}
\end{algorithm}

\textbf{Matching.}
The scoring stage produces a ranked list of DNN data \textit{pages} and, for each page, a ranked list of candidate DRAM \textit{pages}. The matching stage now walks through these lists to build the actual DNN-to-DRAM mapping. An exhaustive search over all $N$ vulnerable DRAM \textit{pages} in $\mathcal{P}_{\text{DRAM}}$ for each of the $M$ DNN data \textit{pages} is computationally prohibitive. Instead, we use the ranking vector $\mathbf{r}_p$ to restrict the search to the top-$K$ candidate DRAM \textit{pages} (typically $K=20$), reducing the number of evaluations per DNN data \textit{page} from $N$ to $K$.

For each candidate DRAM \textit{page} $d$ in the top-$K$ set, we apply all of its profiled bit-flips to DNN data \textit{page} $p$ and evaluate the resulting model through standard inference. This empirical validation is necessary to capture the nonlinear interactions between multiple bit-flips and their combined effect on DNN outputs as the static scores from the previous stage cannot do so. Among the tested candidates, we select the DRAM \textit{page} $d^*$ that yields the largest ASR improvement while satisfying the accuracy constraint $\text{TA} \geq \alpha_{\min}$. The mapping $(p, d^*)$ is recorded and $d^*$ is added to the set of used DRAM \textit{pages} $\mathcal{U}$ to prevent future assignment conflicts. The process then advances to the next highest-ranked DNN data \textit{page} and continues until either the target ASR $\tau$ is achieved or all DNN data \textit{pages} are exhausted. If none of the top-$K$ candidates improve the ASR for a given page, that page is mapped to a DRAM \textit{page} with no bit-flips to preserve model functionality.

The key assumption behind the top-$K$ restriction is that DRAM \textit{pages} with lower vulnerability scores contribute negligibly to backdoor efficacy, allowing us to safely prune them from the search. This transforms an intractable combinatorial optimization into an efficient greedy search with constant cost per DNN data \textit{page}, as formalized in Algorithm~\ref{alg:dram_matching}. Crucially, because the matching evaluates all bit-flips on each candidate DRAM \textit{page}, instead of only specific target bit-flips as in prior methods, ROBBIN constructs a backdoored DNN model whose efficacy accounts for bit-flips achieved via Rowhammer on a given device.

%\vspace{-0.25cm}
\subsubsection{Search Efficiency of DRAM Page Matching}\label{subsubsec:optimality}
The matching problem is inherently challenging because the attacker must assign $M$ DNN data \textit{pages} to a pool of $N$ vulnerable DRAM \textit{pages}, and the backdoor effect of each assignment depends on which other pages have already been mapped. A brute-force search over all possible assignments is clearly infeasible: even for our experimental setup with $M{=}265$ DNN data \textit{pages} and $N{\geq}18k$ vulnerable DRAM \textit{pages}, the number of candidate mappings is prohibitively large. A simpler greedy strategy that tests every DRAM \textit{page} for each DNN data \textit{page} would require $M {\times} N$ model evaluations, which is still expensive when $N$ is large and may result in suboptimal result.

Algorithm~\ref{alg:dram_matching} makes this search tractable in two ways. The scoring step (Eq.~\ref{eq:imp-score1} and~\ref{eq:imp-score2}) precomputes a ranking of DRAM \textit{page} candidates for each DNN data \textit{page} using lightweight arithmetic over the vulnerability matrices, without any model inference. The matching step then evaluates only the top-$K$ candidates (typically $K{=}20$) through actual model inference, reducing the total number of forward passes to at most $M {\times} K$.

This greedy approach is effective because the matching problem empirically exhibits diminishing returns: the first few page assignments contribute the largest ASR gains, while later assignments add progressively less. Our experimental results (Figure~\ref{fig:anoFP32}) confirm this: ASR rises steeply with the first 10--15 DRAM \textit{pages} and saturates well before all DNN data \textit{pages} are exhausted, reaching 90\% ASR with only 33--46 pages out of 265 available. This rapid saturation confirms that the scoring function effectively prioritizes the most impactful assignments and that the greedy ordering captures the bulk of the achievable backdoor effect within a small fraction of the total search space.

\vspace{-0.25cm}
\subsection{Attack Deployment}\label{subsec:deployment}

The deployment stage executes the physical attack using the DNN-to-DRAM \textit{page} mappings $\mathcal{M}_P$ produced by Algorithm~\ref{alg:dram_matching}. The adversary first performs memory massaging~\cite{razavi2016flipfengshui} to ensure that the target DNN data \textit{pages} are placed onto the chosen vulnerable DRAM \textit{pages}. Once the prescribed placement is achieved, the adversary executes Rowhammer using the profiled aggressor patterns to inject the backdoor bit-flips into the model weights. Since sustained hammering produces reproducible bit-flips, as observed in prior studies~\cite{jattke2022blacksmith,gerlach2023rowhammer}, the offline fault map reliably predicts the bit-flips induced during the attack. In typical inference deployments, the DNN model is loaded once and then serves many requests, so hammering constitutes a one-time cost amortized over the entire service. The resulting bit-flips persist throughout the session, corrupting every subsequent inference. Section~\ref{subsec:hw-realization} details the concrete implementation of this procedure on our test platform.

%% file: Sections/4.tex
\vspace{-0.25cm}
\section{Experimental Results}\label{sec:Expt}
In this section, we first describe our experimental setup, characterizing the DRAM vulnerability profiles of the devices under test, and then present an end-to-end attack performance analysis that offers key insights into how quantization and device-specific vulnerabilities influence the backdoor attack.

\vspace{-0.25cm}
\subsection{Experimental Setup and Metrics}\label{subsec:setup}
We conduct evaluation on a commodity desktop system equipped with an Intel Core i7-8700 processor, running 64-bit Ubuntu 22.04 with Linux kernel 6.8. To capture device-specific DRAM vulnerabilities, we test three 8GB DDR4 Dual In-line Memory Modules (DIMMs) from leading memory manufacturers: 
\begin{enumerate}[noitemsep,topsep=5pt,leftmargin=15pt]
    \item \textbf{Device A}: SK Hynix HMA81GU6JJR8N,
    \item \textbf{Device B}: Micron Ballistix BLS8G4D240FSB, 
    \item \textbf{Device C}: Samsung M391A1K43BB2-CTD.
\end{enumerate}
For the target DNN models deployed on this hardware, we train ResNet-20 and VGG-16 on the CIFAR-10 dataset in both FP32 and INT8-quantized (using Quantization-Aware Training) formats. ResNet-20 achieves baseline accuracies of 90.21\% (FP32) and 90.73\% (INT8), while VGG-16 achieves 93.34\% (FP32) and 93.56\% (INT8), respectively. We use the following metrics:

\textbf{Rowhammer Metrics evaluating attack complexity:}
\begin{enumerate}[noitemsep,topsep=5pt,leftmargin=15pt]
    \item \textit{Number of bit-flips} ($N_{\text{flip}}$): total physical bit-flips induced by hammering, computed as $N_{\text{flip}} = \sum_{i} D(p_{\text{clean}}^i, p_{\text{hammered}}^i)$ where $D(\cdot,\cdot)$ is the Hamming distance, and
    \item \textit{Number of pages} ($N_{\text{page}}$): total distinct DRAM \textit{pages} requiring hammering, directly affecting the attack execution time.
\end{enumerate}

\textbf{Backdoor metrics evaluating ML backdoor:}
\begin{enumerate}[noitemsep,topsep=5pt,leftmargin=15pt]
    \item \textit{Test Accuracy} (TA): model's classification accuracy on clean test data, which must remain high to avoid detection, and
    \item \textit{Attack Success Rate} (ASR): percentage of triggered test inputs misclassified to the target class, validating successful backdoor.
\end{enumerate}

\vspace{-0.25cm}
\subsubsection{DRAM Vulnerability Profiling}\label{sec:profile}

As described in Section~\ref{sec:threat}, ROBBIN relies on a device-specific fault map obtained through offline profiling. Using Blacksmith~\cite{jattke2022blacksmith}, we characterize the three DRAM devices to obtain deterministic vulnerability patterns.

\begin{figure*}[t]
    \centering
    \begin{minipage}[b]{0.32\textwidth}
        \centering
        \includegraphics[width=0.85\linewidth, trim=0 0 0 0, clip]{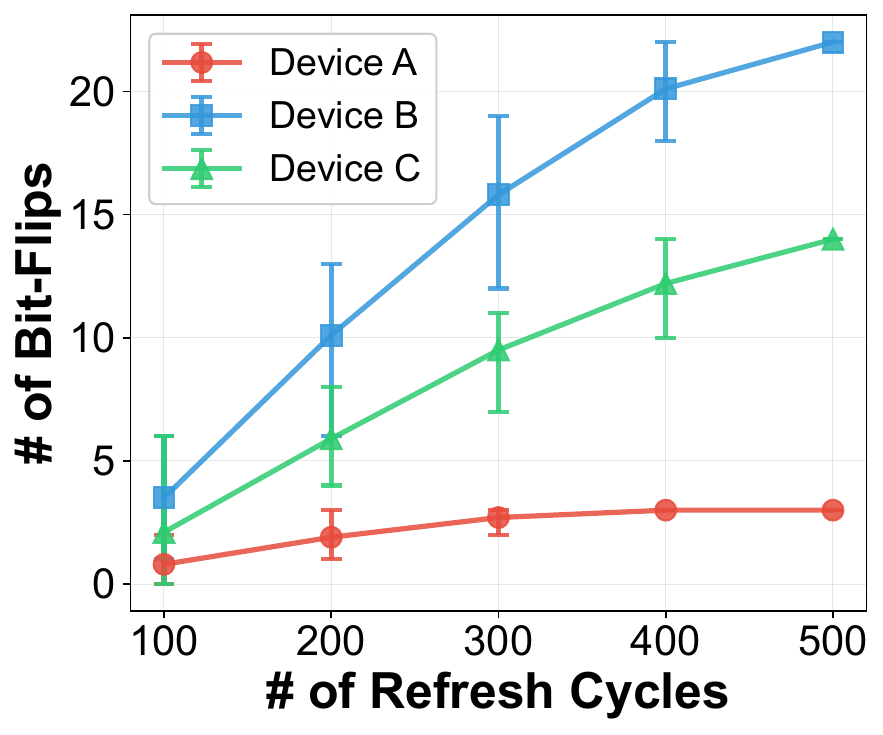}
        \caption{DRAM vulnerability profiling across three devices showing bit-flip range for one victim row versus refresh cycles under 4-row hammering.}
        \label{fig:Hist-profile}
    \end{minipage}
    \hfill
    \begin{minipage}[b]{0.65\textwidth}
        \centering
        \begin{subfigure}[b]{0.48\linewidth}
            \centering
            \includegraphics[width=0.95\linewidth, trim=0 0 0 0, clip]{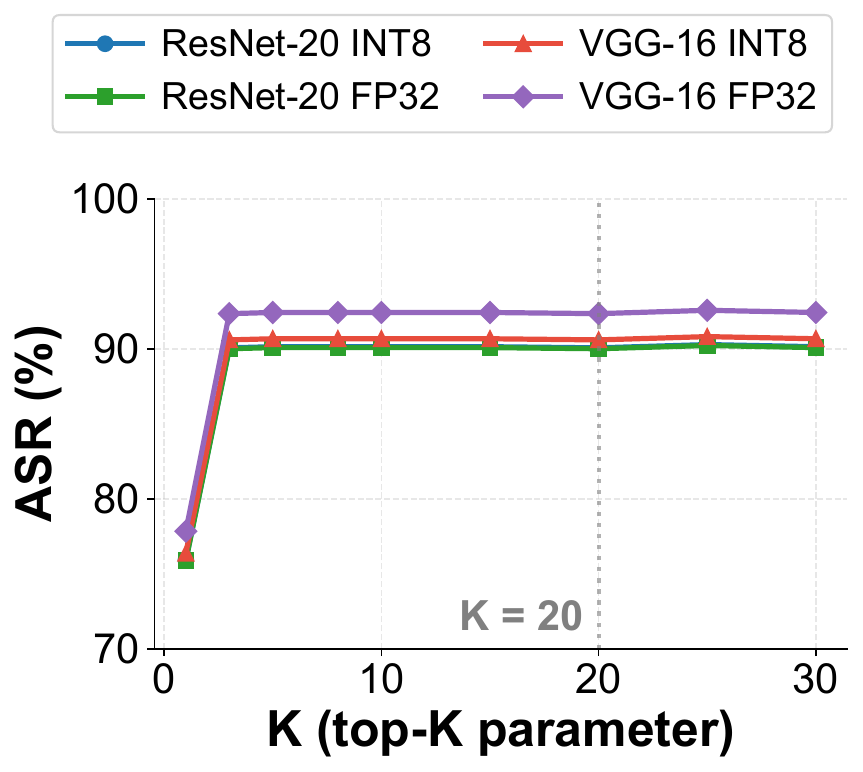}
            \caption{ASR vs. top-$K$.}
            \label{fig:topK-ASR}
        \end{subfigure}
        \hfill
        \begin{subfigure}[b]{0.48\linewidth}
            \centering
            \includegraphics[width=0.92\linewidth, trim=0 0 0 0, clip]{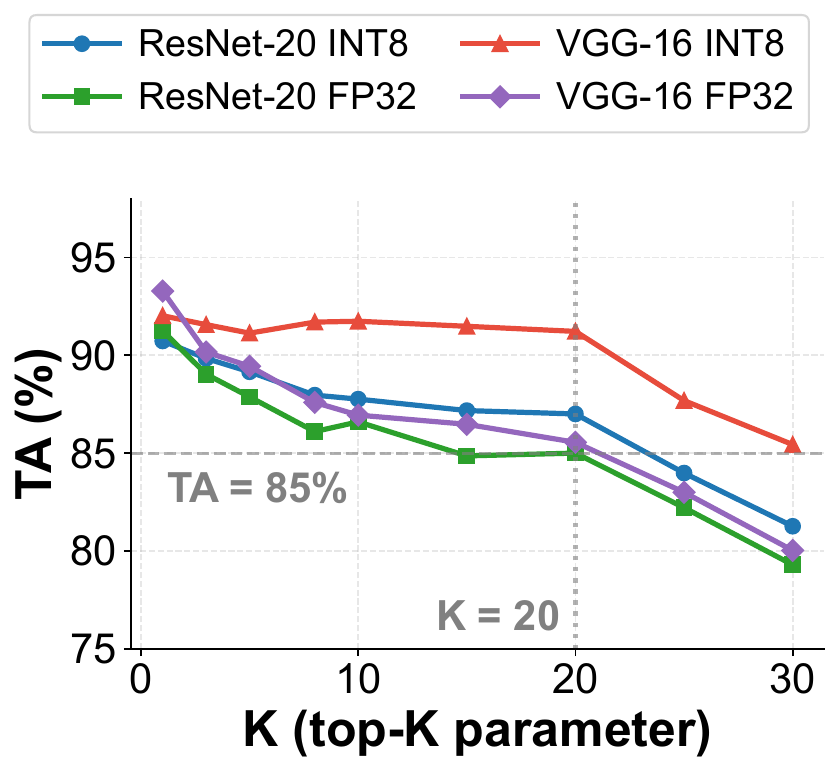}
            \caption{TA vs. top-$K$.}
            \label{fig:topK-TA}
        \end{subfigure}
        %\vspace{-0.25cm}
        \caption{Effect of the top-$K$ candidate pool size on attack performance across different networks: (a) ASR and (b) TA on Device B.}
        \label{fig:param-sensitivity}
    \end{minipage}
    %\vspace{-0.25cm}
\end{figure*}

Figure~\ref{fig:Hist-profile} shows the number of bit-flips induced in a single DRAM victim row as a function of refresh cycles under 4-row hammering for all three devices. The error bars capture the range of bit-flips observed across 10 repetitions, illustrating how the bit-flip behavior transitions from probabilistic to deterministic. At lower refresh cycles (100--200), the variation is substantial, but all devices converge to a stable, reproducible bit-flip count by 500 refresh cycles. The vulnerability density varies substantially across devices: Device A exhibits the fewest bit-flips per page across 18,637 vulnerable pages (7.1\% of 262k total pages), Device B shows moderate vulnerability across 96,293 pages (36.7\%), and Device C is the most vulnerable with 159,719 pages (60.9\%). While higher hammering intensities, such as 7-row and 15-row patterns, increase the number of bit-flips per page, we adopt 4-row hammering throughout our experiments as it provides sufficient bit-flips for backdoor construction while minimizing collateral damage. This saturation at higher refresh cycle count enables us to use Rowhammer as a predictable fault-injection mechanism, thereby facilitating the construction of a reliable backdoor attack across multiple commodity DRAM chips.

\vspace{-0.25cm}
\subsection{End-to-End Performance Comparison}

We demonstrate practical backdoor attacks on the three profiled DRAM devices using both FP32 and INT8 models for ResNet-20 and VGG-16 on CIFAR-10. In all cases, we use a learned 10 x 10 patch trigger placed at the bottom-right corner, and target class 2. The ROBBIN codebase is publicly available at \url{https://github.com/noisyor/ROBBIN-Rowhammer-Based-Backdoor-Injection-during-Inference}.

\vspace{-0.25cm}
\subsubsection{Attack Parameters}
Each targeted DRAM \textit{page} undergoes 500 refresh cycles of hammering to induce the deterministic bit-flip patterns identified in the device fault map (Section~\ref{sec:profile}). %Based on Figure~\ref{fig:Hist-profile}, we employ 4-row hammering intensity across all experiments to minimize the average number of bit-flips per page. 

With this setting, $K$ \textit{pages} are selected from a pool of $N\geq18K$ vulnerable DRAM \textit{pages} across devices, where ResNet-20 occupies $M=265$ (FP32) or $66$ (INT8) DNN data \textit{pages} and VGG-16 occupies $M=14{,}905$ (FP32) or $3{,}657$ (INT8) \textit{pages}. We select the $K$ parameter using a validation rule. Since larger $K$ increases candidate coverage but also introduces lower-ranked DRAM \textit{pages} that can amplify damage to test accuracy, we choose $K$ as the largest value that preserves the test accuracy constraint $\alpha_{\min}$. As shown in Fig.~\ref{fig:param-sensitivity}(a), ASR saturates at smaller $K$ for all four configurations spanning ResNet-20 and VGG-16 in both INT8 and FP32, while Fig.~\ref{fig:param-sensitivity}(b) shows that TA remains above 85\% up to $K=20$ across the same four curves and degrades beyond this point. \begin{wrapfigure}[16]{l}{0.23\textwidth}
\vspace{-0.45cm}
  \begin{center}
    \includegraphics[scale=0.32, trim=3 0 0 0, clip]{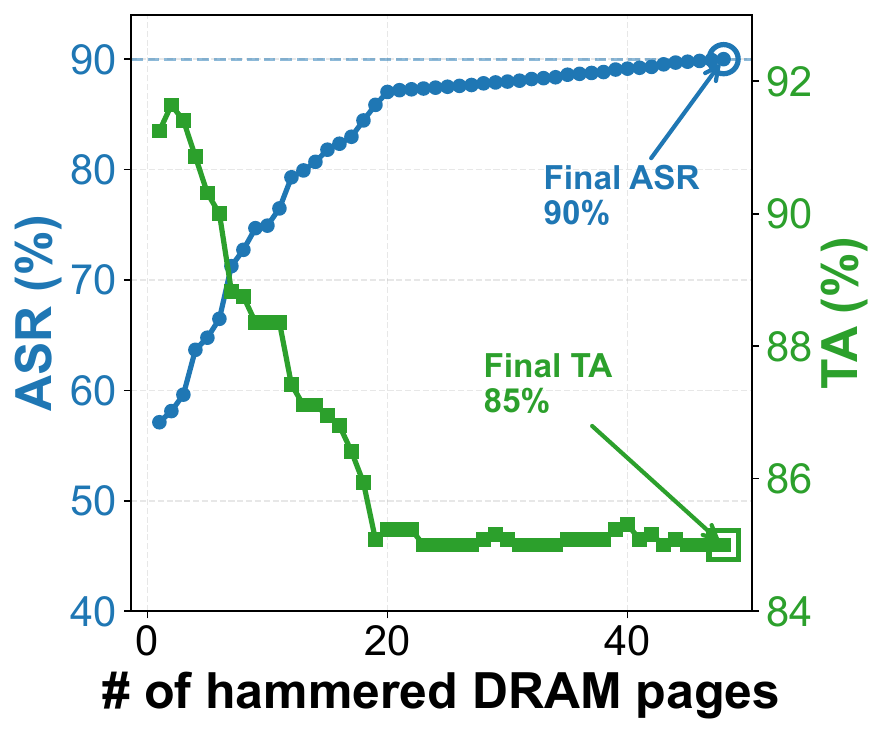}
    \vspace{-0.5cm}
    \caption{ROBBIN's progression vs. \# of hammered DRAM \textit{pages} at $K=20$.}
    \label{fig:anoFP32}
   \end{center}
\vspace{-0.05cm}
\end{wrapfigure} In our experiments, we find that $K=20$ is a suitable parameter for ROBBIN across architectures and quantization schemes, and we use it for all subsequent experiments because it provides broad search coverage without incurring the accuracy loss observed at larger $K$. With $K=20$ fixed, Fig.~\ref{fig:anoFP32} illustrates how, for ROBBIN on Device~B with FP32 quantization, ASR gradually rises to 90\% as the number of hammered DRAM \textit{pages} increases while meeting the TA constraint, confirming the effectiveness of Algorithm~\ref{alg:dram_matching}. The same phenomenon is observed on other DRAM devices and DNN models, where the mapping achieved by Algorithm~\ref{alg:dram_matching} leads to a reliable and successful backdoor attack.

To assess whether the proposed scoring function and top-$K$ greedy matching preserve the attainable backdoor effect, we evaluate ROBBIN against three internal variants: a sensitivity-only greedy baseline that removes the hardware-aware ranking vector $\mathbf{r}_p$, a target-only variant that ignores collateral bit-flips during matching, and a random DNN-to-DRAM \textit{page} assignment that serves as a lower bound. ROBBIN achieves high ASR together with high TA, meeting both objectives of the attack, while the three baselines cannot reach $> 50\%$ ASR without sacrificing TA, and their ASR remains well below ROBBIN's even after the TA budget is relaxed. This contrast shows that the ranking vector prunes low-value candidates effectively and that accounting for all profiled bit-flips during matching is necessary for a reliable backdoor construction.

Having established that ROBBIN is well calibrated with respect to its own design choices, we next turn to external comparisons with prior bit-flip backdoor attacks, namely Don't Knock~\cite{tol2023don} designed for INT8 and OneFlip~\cite{li2025oneflip} which targeted FP32. To keep these comparisons fair and isolate the contribution of the matching strategy itself, we reuse the same profiling results across all methods and hold the backdoor attack parameters, including trigger size and trigger location, identical throughout.

\vspace{-0.25cm}
\subsubsection{INT8 Performance: Comparison with Don't Knock}
\begin{figure}[t]
    \centering
    \begin{subfigure}[t]{\linewidth}
        \centering
        \includegraphics[width=0.9\linewidth, trim=0 10 0 0, clip]{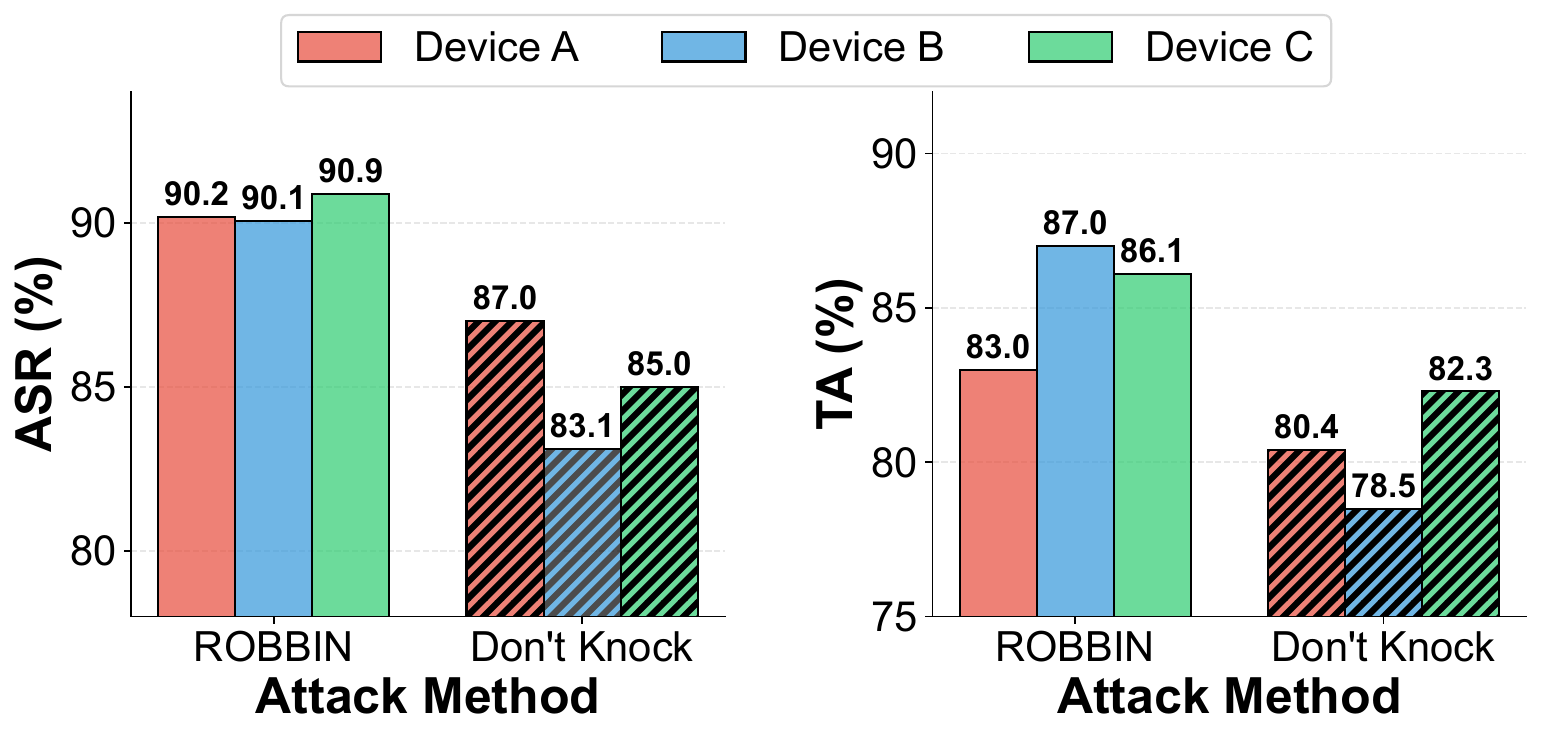}
        \caption{ResNet-20 on CIFAR-10}
        \label{fig:int8-resnet}
    \end{subfigure}
    \vspace{0.1cm}
    \begin{subfigure}[t]{\linewidth}
        \centering
        \includegraphics[width=0.9\linewidth, trim=0 10 0 0, clip]{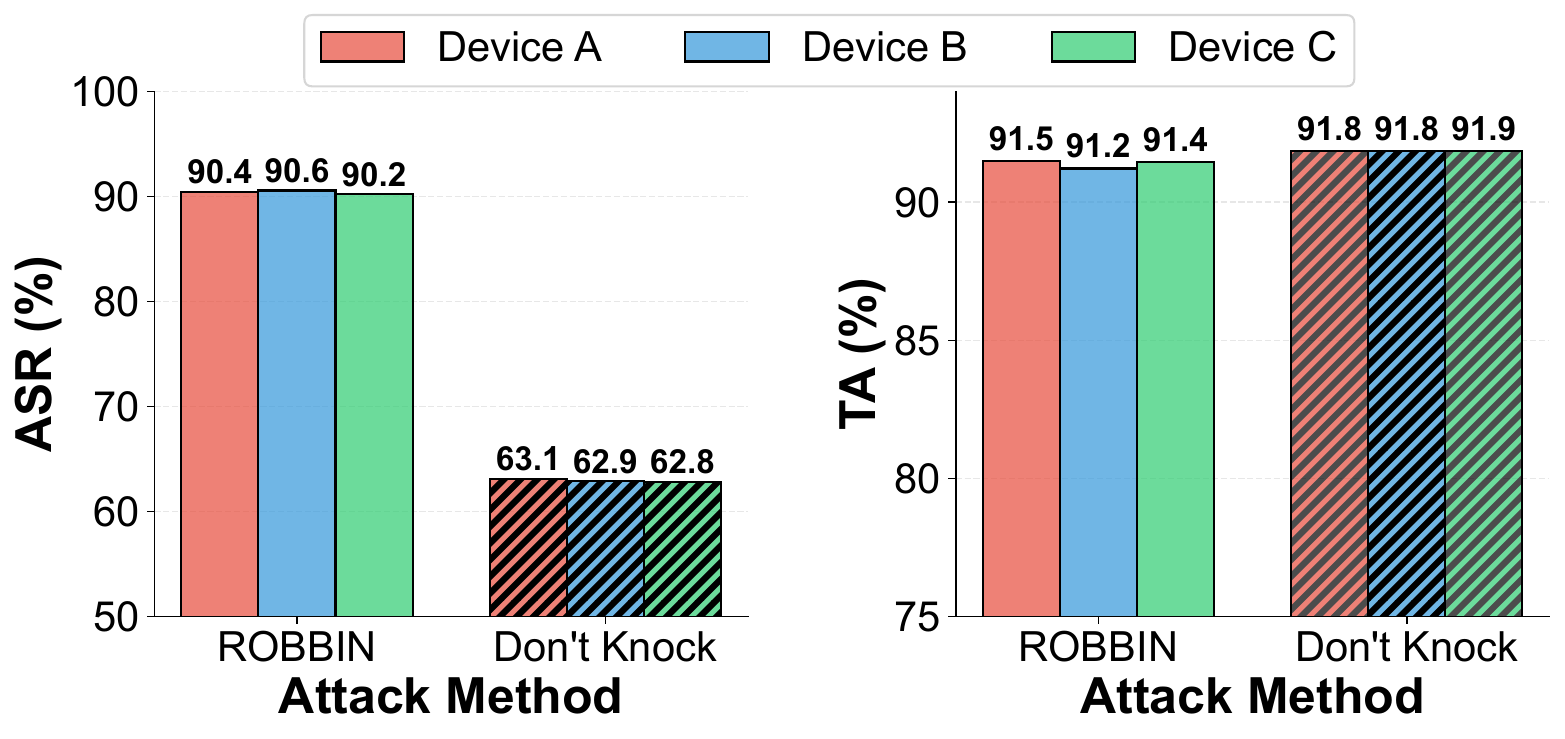}
        \caption{VGG-16 on CIFAR-10}
        \label{fig:int8-vgg}
    \end{subfigure}
    \vspace{-0.15cm}
    \caption{Comparing ROBBIN with Don't Knock~\cite{tol2023don} for \textbf{INT8 quantization} across three hardware devices.}
    \label{fig:int8-metric}
    \vspace{-0.25cm}
\end{figure}

Figure~\ref{fig:int8-metric} compares the ASR and TA of ROBBIN against Don't Knock~\cite{tol2023don} for INT8 models across all three devices. For ResNet-20 (Fig.~\ref{fig:int8-resnet}), ROBBIN achieves ASR of approximately 90\% across all devices, outperforming Don't Knock, which achieves ASR between 83.1\% and 87.0\%. The average TA for ROBBIN is 85.4\%, compared to 80.4\% for Don't Knock. For VGG-16 (Fig.~\ref{fig:int8-vgg}), ROBBIN maintains a consistent ASR above 90\%, while Don't Knock achieves only 62.8\% to 63.1\% ASR across all devices. Despite this large gap in ASR, the TA for both methods remains comparable on VGG-16, averaging around 91.4\% for ROBBIN and 91.8\% for Don't Knock, indicating that it preserves model accuracy on clean inputs but fails to reliably trigger the backdoor.

Table~\ref{tab:int8-complexity} provides the detailed attack complexity comparison. For ResNet-20, ROBBIN requires 64\% to 73\% fewer DRAM \textit{pages} and 25\% to 30\% fewer total bit-flips than Don't Knock across all devices. For VGG-16, ROBBIN achieves even larger reductions on Devices B and C, requiring approximately 79\% fewer pages and 87\% to 89\% fewer bit-flips, while Device A requires 48 pages compared to Don't Knock's 95, with 63\% fewer bit-flips. Don't Knock's lower TA on ResNet-20 stems from its decoupled backdoor construction, which limits it to a single bit-flip per DNN data \textit{page}, whereas the actual hardware implementation includes unwanted collateral bit-flips. In contrast, ROBBIN's DRAM \textit{page} selection incorporates all available bit-flips during backdoor construction, thereby achieving both backdoor objectives across multiple DRAMs.

\begin{tcolorbox}[colback=gray!10, colframe=black, boxrule=0.5pt, arc=4pt, left=1pt, right=1pt, top=1pt, bottom=1pt]
\centering
\textit{\textbf{Key Takeaway:} ROBBIN achieves consistently high ASR while requiring substantially fewer DRAM \textit{pages} than Don't Knock~\cite{tol2023don}.}
\end{tcolorbox}

\begin{table}[t]
\centering
\caption{Attack complexity comparison between ROBBIN and Don't Knock~\cite{tol2023don} for INT8 quantization across three DRAMs.}
\label{tab:int8-complexity}
\vspace{-0.15cm}
\footnotesize
\begin{tabular}{llcccccc}
\toprule
& & \multicolumn{2}{c}{\textbf{Device A}} & \multicolumn{2}{c}{\textbf{Device B}} & \multicolumn{2}{c}{\textbf{Device C}} \\
\cmidrule(lr){3-4} \cmidrule(lr){5-6} \cmidrule(lr){7-8}
\textbf{Model} & \textbf{Method} & $N_{\text{page}}$ & $N_{\text{flip}}$ & $N_{\text{page}}$ & $N_{\text{flip}}$ & $N_{\text{page}}$ & $N_{\text{flip}}$ \\
\midrule
\multirow{2}{*}{ResNet-20} & ROBBIN & 16 & 70 & 14 & 86 & 13 & 79 \\
& Don't Knock & 45 & 100 & 50 & 115 & 48 & 108 \\
\midrule
\multirow{2}{*}{VGG-16} & ROBBIN & 48 & 201 & 23 & 391 & 22 & 434 \\
& Don't Knock & 95 & 539 & 112 & 3409 & 105 & 3341 \\
\bottomrule
\end{tabular}
\vspace{-0.4cm}
\end{table}

\vspace{-0.25cm}
\subsubsection{FP32 Performance: Comparison with OneFlip}

\begin{figure}[t]
    \centering
    \begin{subfigure}[t]{\linewidth}
        \centering
        \includegraphics[width=0.9\linewidth, trim=0 10 0 0, clip]{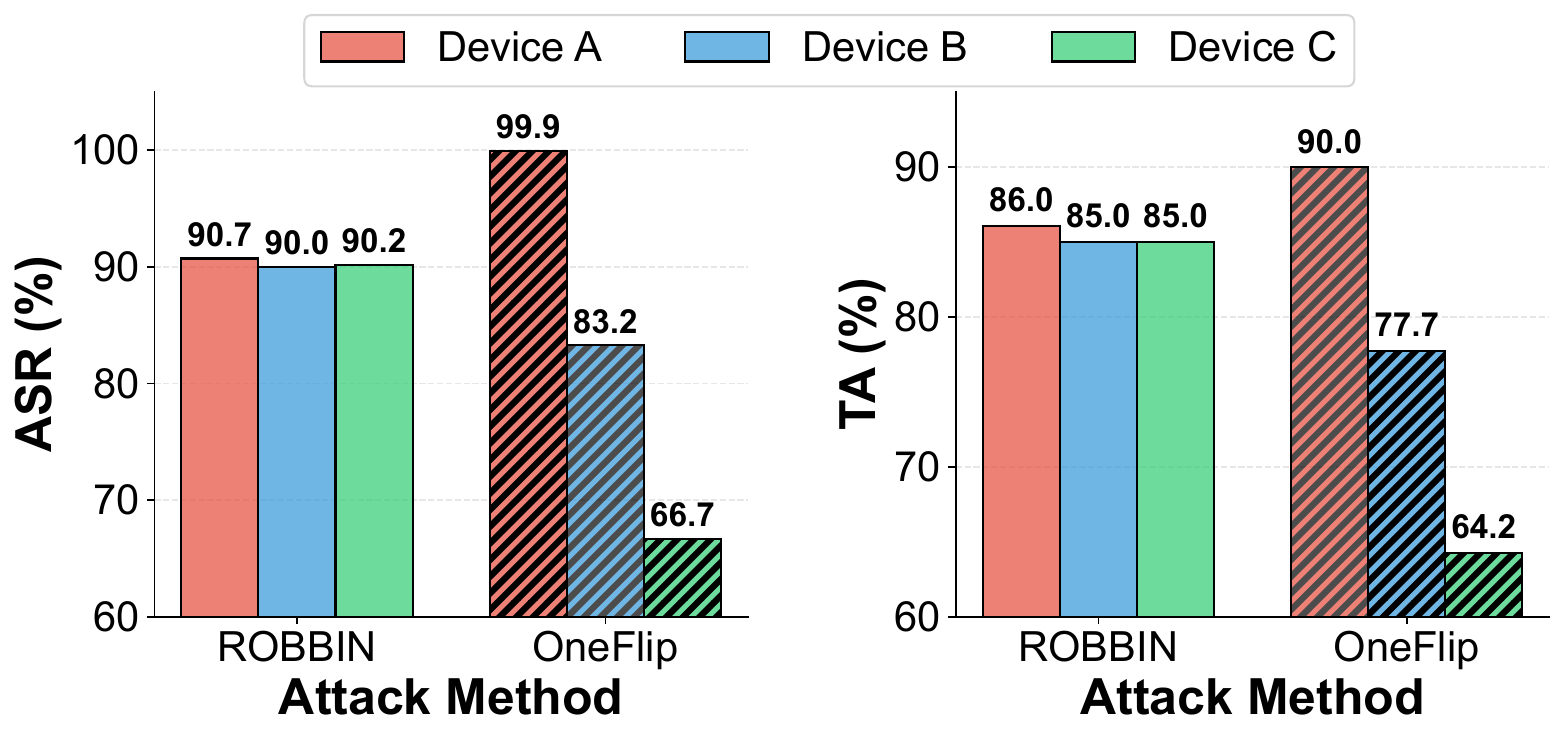}
        \caption{ResNet-20 on CIFAR-10}
        \label{fig:fp32-resnet}
    \end{subfigure}
    \vspace{0.1cm}
    \begin{subfigure}[t]{\linewidth}
        \centering
        \includegraphics[width=0.9\linewidth, trim=0 10 0 0, clip]{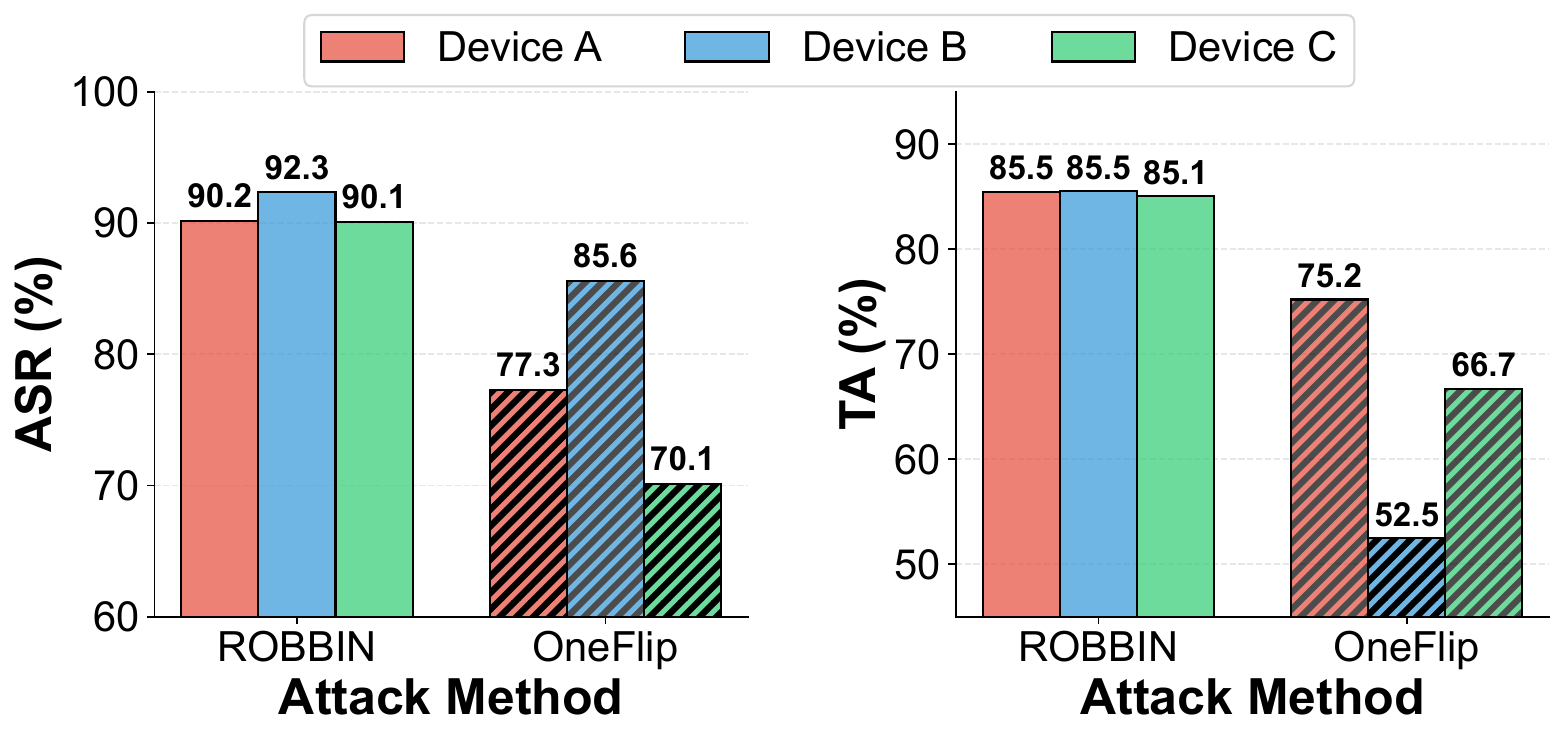}
        \caption{VGG-16 on CIFAR-10}
        \label{fig:fp32-vgg}
    \end{subfigure}
    \vspace{-0.15cm}
    \caption{Comparing ROBBIN with OneFlip~\cite{li2025oneflip} for \textbf{FP32 quantization} across three hardware devices.}
    \label{fig:fp32-metric}
    \vspace{-0.5cm}
\end{figure}

Figure~\ref{fig:fp32-metric} compares the ASR and TA of ROBBIN against OneFlip~\cite{li2025oneflip} for FP32 models. For ResNet-20 (Fig.~\ref{fig:fp32-resnet}), ROBBIN achieves a consistent ASR of $\approx$90\% across all devices while maintaining TA close to 85.0\%. OneFlip's performance, however, varies substantially across devices. While it performs well on Device A, where the low vulnerability density limits collateral damage, its effectiveness degrades on devices with denser bit-flip patterns: on Device B, ASR drops to 83.2\% with 77.7\% TA, and on Device C, it falls further to 66.7\% ASR with only 64.2\% TA.

This degradation becomes even more pronounced for VGG-16 (Fig.~\ref{fig:fp32-vgg}), where collateral bit-flips are particularly damaging because each DRAM \textit{page} contains more weight parameters from the larger model. ROBBIN maintains its consistent performance with ASR above 90\% and TA above 85\% across all devices. In contrast, OneFlip's TA collapses to 75.2\%, 52.5\%, and 66.7\% on Devices A, B, and C, respectively, with ASR also degrading to 77.3\%, 85.6\%, and 70.1\%. The fundamental issue is that collateral bit-flips landing in highly sensitive regions of the FP32 representation, such as the exponent or sign bit, cause disproportionate damage to model accuracy that OneFlip cannot account for.

ROBBIN avoids this fragility by distributing the backdoor across multiple DRAM \textit{pages}, where the collateral bit-flips on each page are explicitly accounted for during the DRAM \textit{page} matching process (Algorithm~\ref{alg:dram_matching}). By incorporating all available bit-flips during backdoor construction rather than treating them as unwanted noise, ROBBIN achieves reliable ASR close to 90\% with TA above 85\% across all DRAMs and both model architectures.

\begin{tcolorbox}[colback=gray!10, colframe=black, boxrule=0.5pt, arc=4pt, left=1pt, right=1pt, top=1pt, bottom=1pt]
\centering
\textit{\textbf{Key Takeaway:} ROBBIN maintains consistent ASR and TA across all devices, while OneFlip~\cite{li2025oneflip} suffers from unpredictable collateral damage that severely degrades TA, particularly for larger models.}
\end{tcolorbox}

\vspace{-0.25cm}
\subsubsection{Hardware Realization of ROBBIN}\label{subsec:hw-realization}
All results reported above are obtained from an end-to-end hardware realization on the Intel Core i7-8700 (Coffee Lake) platform. We realize the DNN-to-DRAM \textit{page} mapping $\mathcal{M}_P$ produced by Algorithm~\ref{alg:dram_matching} through memory massaging that exploits the Linux kernel's Per-CPU Pageset (PCP) LIFO allocation policy~\cite{razavi2016flipfengshui}. Specifically, we first evict the target DNN data \textit{pages} from physical memory using \texttt{madvise}. We drain the OS buddy allocator and selectively free the chosen vulnerable DRAM \textit{pages} to the top of the PCP list. When the model weights are re-accessed during inference, the resulting page faults place the target DNN \textit{pages} onto the intended vulnerable DRAM \textit{pages}. Once placement is complete, Rowhammer fault injection replays the profiled aggressor patterns. The dominant setup cost is the one-time construction of the device-specific fault map, whereas ROBBIN's recurring deployment cost is governed by page placement and hammering of only the selected DRAM \textit{pages}. Across our experiments, ROBBIN hammers only 18--130 DRAM \textit{pages} to implement a reliable backdoor. Once injected, the resulting corruption persists for the duration of the loaded inference session.

%All results reported above are obtained from an end-to-end hardware realization on the Intel Core i7-8700 (Coffee Lake) platform. We realize the DNN-to-DRAM \textit{page} mapping $\mathcal{M}_P$ produced by Algorithm~\ref{alg:dram_matching} through memory massaging that exploits the Linux kernel's Per-CPU Pageset (PCP) LIFO allocation policy~\cite{razavi2016flipfengshui}. Specifically, we first evict the target DNN data \textit{pages} from physical memory using \texttt{madvise}. We then allocate a 7\,GB buffer to drain the OS buddy allocator and selectively free the chosen vulnerable DRAM \textit{pages} to the top of the PCP list. When the model weights are re-accessed during inference, the resulting page faults place the target DNN \textit{pages} onto the intended vulnerable DRAM \textit{pages}. Once placement is complete, Rowhammer fault injection replays the profiled aggressor patterns. Across our experiments, ROBBIN hammers only 18--130 DRAM \textit{pages}, so the one-time injection cost scales with the number of selected pages rather than with an exhaustive search over all vulnerable pages. The induced bit-flips then persist for the entire inference session until the model is loaded again.

%% file: Sections/5.tex
\vspace{-0.25cm}
\section{Practical Security Implications}\label{sec:sec-implication}
Having seen in Section~\ref{sec:Expt} that ROBBIN achieves reliable cross-device backdoor injection, we now turn to an examination of its scalability and detectability, followed by a system-level countermeasure to prevent such attacks.
\vspace{-0.25cm}

\subsection{Scalability to Larger Datasets and Models}
\label{sec:scalability}

To evaluate ROBBIN's scalability, we apply it to an INT8-quantized ResNet-50 on CIFAR-10 and ImageNet, extending the evaluation to a deeper architecture and from 10 to 1000 classes. On CIFAR-10, ROBBIN achieves an ASR of $84.4\%$, while clean accuracy decreases from $90.9\%$ to $87.3\%$ (a $3.6\%$ drop). On ImageNet, it achieves a $97.7\%$ ASR, while clean accuracy decreases from $88.8\%$ to $84.9\%$ (a $3.9\%$ drop). For ImageNet, we keep the trigger area at approximately $10\%$ of the input image, consistent with the trigger fraction used for CIFAR-10 in prior work~\cite{chen2021proflip}. These results demonstrate that ROBBIN's scoring and matching procedure remains effective when scaling to both deeper models and substantially larger label spaces. Extending this evaluation to larger architectures and other model families, particularly transformer-based vision and language models, where backdoor attacks remain an emerging research area~\cite{li2025backdoorllm}, is a natural direction for future work.

% \subsection{Scalability to Larger Datasets and Models}
% \label{sec:scalability}

% To evaluate the scalability of ROBBIN, we target an INT8 quantized ResNet50 on CIFAR 10 and ImageNet, thereby moving to both a deeper architecture and a substantially larger label space. On ImageNet, the trigger area is kept at approximately 10\% of the input image, consistent with the trigger fraction used for CIFAR 10 in prior work~\cite{chen2021proflip}. Table~\ref{tab:scalability} shows that ROBBIN attains an ASR of $84.4\%$ on CIFAR 10 and $97.7\%$ on ImageNet, with only a small drop in clean accuracy. These results show that the scoring and matching procedure remains effective not only for deeper models, but also when scaling from a 10 class to a 1000 class setting. Extending this study to even larger models and to other architectural families, in particular transformer-based vision and language models where backdoor attacks remain an emerging and largely open area of research~\cite{li2025backdoorllm}, is a natural direction for future work.

% \begin{table}[t]
% \centering
% \caption{ROBBIN scalability results on ResNet-50 (INT8).}
% \label{tab:scalability}
% \small
% \vspace{-0.25cm}
% \begin{tabular}{@{}llrrr@{}}
% \toprule
% Model & Dataset & \shortstack{Clean Acc.\\ Before} & \shortstack{Clean Acc.\\ After} & ASR \\
% \midrule
% ResNet-50 & CIFAR-10 & 90.9\% & 87.3\% & 84.4\% \\
% ResNet-50 & ImageNet & 88.8\% & 84.9\% & 97.7\% \\
% \bottomrule
% \end{tabular}
% \vspace{-0.25cm}
% \end{table}
% \vspace{-0.25cm}
\subsection{Detectability}\label{subsec:detect}
ROBBIN's backdoor is transient: the bit-flips exist only in the DRAM-resident copy of the model during the active inference session and are never written back to persistent storage. When the model is reloaded from disk, whether for routine maintenance, periodic refresh, or explicit security audit, the weights revert to their clean state, and all evidence of the backdoor disappears.
 
The transient nature of ROBBIN can evade audit procedures that reload the model from disk before analysis, because reloading reinstantiates a clean model. Neural Cleanse~\cite{wang2019neural}, STRIP~\cite{gao2019strip}, and PSBD~\cite{li2025psbd} all operate on a loaded model instance, so any audit procedure that reloads the model from disk for inspection instantiates a clean copy with no trace of the backdoor for these defenses to act on. Under continuous runtime monitoring, however, Algorithm~\ref{alg:dram_matching} constrains overall TA while focusing on aggregate model behavior rather than per-class or detector-specific characteristics. Evaluating detectability under live monitoring, including classwise activation based defenses such as Activation Clustering~\cite{chen2019detecting}, remains an important line of investigation.

\begin{figure}[t]
    \centering
    \begin{subfigure}[t]{0.45\linewidth}
        \centering
        \includegraphics[width=\linewidth]{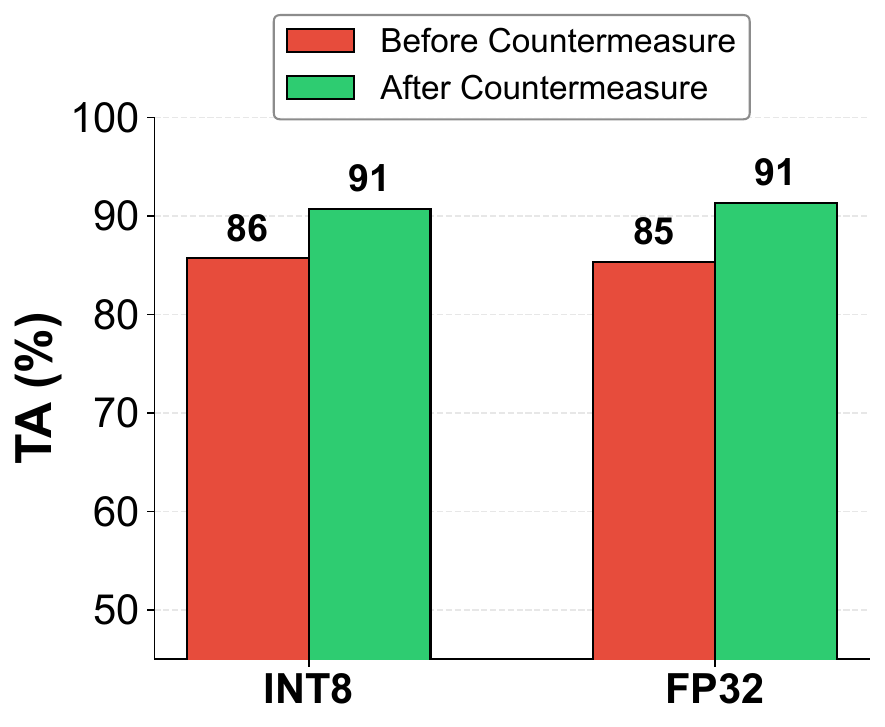}
        \caption{Test Accuracy (TA)}
        \label{fig:counter-ta}
    \end{subfigure}
    \hfill
    \begin{subfigure}[t]{0.45\linewidth}
        \centering
        \includegraphics[width=\linewidth]{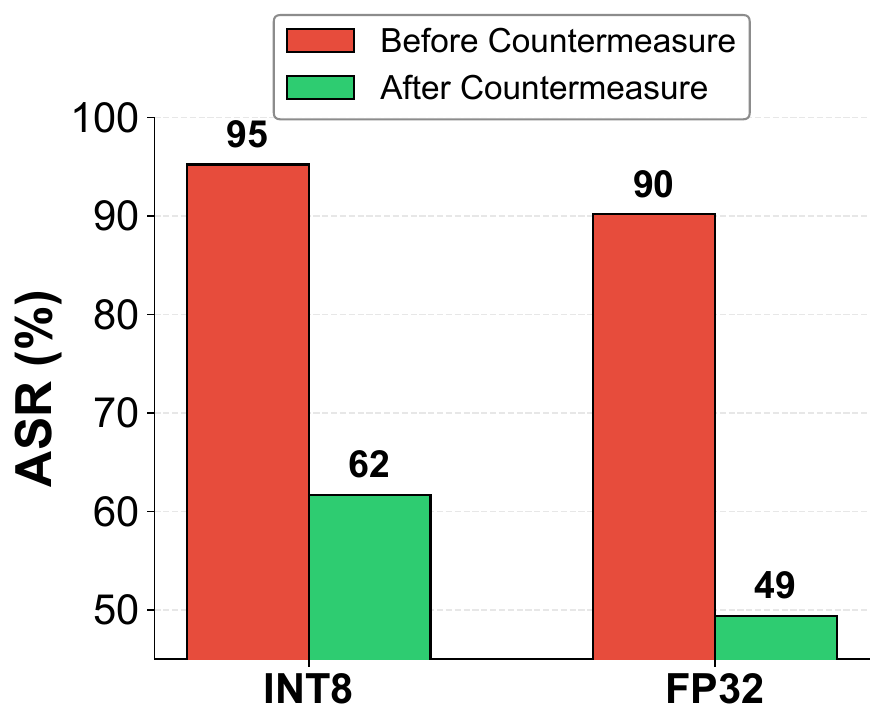}
        \caption{Attack Success Rate (ASR)}
        \label{fig:counter-asr}
    \end{subfigure}
    \vspace{-0.15cm}
    \caption{Vulnerability-aware page allocation on ResNet-20 as a countermeasure that restores clean TA and reduces ASR.}
    \label{fig:countermeasure}
    \vspace{-0.5cm}
\end{figure}

\vspace{-0.25cm}
\subsection{Proposed Countermeasure}\label{subsec:counter}
ROBBIN relies on placing model data onto physical memory pages associated with reproducible Rowhammer bit flips, which suggests a natural systems-level countermeasure: vulnerability-aware page allocation. Under a complete vulnerability profile, the defender can exclude profiled vulnerable pages from allocation to model weights, leaving Algorithm~\ref{alg:dram_matching} with no exploitable placements. This defense is software mediated and does not require hardware modification, although it does require allocator support to control physical placement during model loading.

Vulnerability-aware page allocation exposes a clear security versus memory trade-off rather than serving as a uniform deployment policy. Its cost depends on the profiled vulnerable page density of the device, which in our study ranges from 7.1\% to 60.9\%. As shown in Fig.~\ref{fig:countermeasure}, blacklisting profiled vulnerable pages on Device~B restores clean accuracy from 86\% to 91\% in INT8 and from 85\% to 91\% in FP32, while reducing ASR from 95\% to 62\% and from 90\% to 49\%, respectively. These reduced ASR values match the trigger only baseline for ResNet 20, defined as the misclassification rate induced by the trigger pattern in the absence of any weight corruption. Together, these results indicate that vulnerable page availability is a major systems factor governing attack efficacy.

Vulnerability-aware blacklisting is most attractive on devices with low vulnerable page density, where it can provide strong protection at modest memory cost. On more vulnerable devices, the same mechanism remains applicable but incurs a larger capacity penalty, which motivates more selective policies. Since ROBBIN achieves high ASR using only 18 to 130 pages, selectively blacklisting the pages with the highest reproducible flip density may provide a lower-cost defense while preserving much of the security benefit. These observations position vulnerability-aware allocation as an attractive building block for Rowhammer-resilient ML deployments, subject to further evaluation against adaptive attackers.

%% file: Sections/6.tex
\vspace{-0.25cm}
\section{Conclusion}\label{sec:conclusion}

In this paper, we propose ROBBIN, a hardware-aware Rowhammer-based inference-time backdoor attack that folds device-specific DRAM vulnerability profiles directly into backdoor construction. Treating bit-flip injection and backdoor design as independent problems makes performance unreliable across DRAM chips. By accounting for the physical characteristics of the target device, ROBBIN turns the irregularities of commodity DRAM into a reliable attack primitive, achieving ASR close to $90\%$ with test accuracy above $83\%$ on every one of three DDR4 devices across two architectures and two quantization formats. The same DRAM profiling can also be repurposed as a defense that guides safe model placement, which shows that hardware-aware attack design is not only more effective but also essential for accurately assessing the real-world risk of Rowhammer-based threats to deployed ML systems.

\vspace{-0.25cm}
\section*{Acknowledgment}
This work was supported in part by the NSF under grant CNS-1916762 and industry partners of NSF IUCRC CHEST.

% In this paper, we propose ROBBIN, a hardware-aware Rowhammer-based inference-time backdoor injection attack that integrates device-specific DRAM vulnerability profiles directly into the backdoor construction process. We identify that treating bit-flip injection and backdoor design as independent problems renders attack performance unreliable across DRAM chips. This is because Rowhammer inherently induces collateral bit flips whose distribution varies across devices, and any backdoor design that ignores this variability suffers from unpredictable degradation in clean accuracy. By incorporating the physical characteristics of the target DRAM device, ROBBIN turns the irregularities of commodity DRAM from a source of noise into a reliable attack primitive.

% Our results across three DDR4 devices, two model architectures, and two quantization formats demonstrate that this hardware-aware approach delivers consistent attack performance where prior methods fail. ROBBIN achieves an ASR close to 90\% with test accuracy above 83\% on every device tested, while Don't Knock~\cite{tol2023don} and OneFlip~\cite{li2025oneflip} degrade unpredictably as collateral bit flips accumulate. We further show that the same DRAM profiling that enables ROBBIN can be turned into a defense, guiding safe model placement to mitigate Rowhammer-based inference-time attacks. Our work thus demonstrates that hardware-aware attack design is not only more effective but also essential for accurately assessing the real-world risk of Rowhammer-based threats to the integrity of deployed ML systems. 